\documentstyle[twoside,fleqn,epsf,espcrc2]{article}

\newcommand{\plus}{\makebox[15pt][c]{$+$}}
\newcommand{\minus}{\makebox[15pt][c]{$-$}}

\newcommand{\er}[2]{\raisebox{0.08em}{\scriptsize {$\;\begin{array}{@{}l@{}}
                          \plus\makebox[0.55em][r]{#1} \\[-0.12em] 
                          \minus\makebox[0.55em][r]{#2} 
                        \end{array}$}}}

\newcommand{\ewxy}[2]{\setlength{\epsfxsize}{#2}\epsfbox[10 60 640 590]{#1}}

\newcommand{\ie} {{\em i.e.}}

\newcommand{\rms}{{\em rms}}

\newcommand{\kappacrit}{$\kappa_{\rm crit}$}
\newcommand{\kappaheavy}{$\kappa_h$}
\newcommand{\kappaheavyprime}{$\kappa_{h'}$}
\newcommand{\kappalight}{$\kappa_l$}
\newcommand{\kappacharm}{$\kappa_{\rm charm}$}
\newcommand{\kappastrange}{$\kappa_s$}

\title{Hopping Parameter Analysis of Leptonic and Semi-Leptonic \\
       Heavy-Light Decays}

\author{UKQCD Collaboration, presented by David Henty,  \\ 
	Department of Physics and Astronomy, The University of Edinburgh,
	Edinburgh EH9 3JZ, Scotland.}

\begin{document}

\begin{abstract}
We study leptonic and semi-leptonic decays of $D$ and $B$ mesons.
Use of the Hopping Parameter Expansion (HPE)
for two-point functions allows us continuously
to vary the pseudoscalar mass from below $m_D$
up towards $m_B$. We compute the pseudoscalar
decay constants $f_D$ and $f_B$, and observe consistency with the value
calculated in the static limit. From the measurement of three-point
functions we compute the matrix element relevant to the decay
$\bar B\to D l \bar\nu_l$
and extract the Isgur-Wise function $\xi(v.v')$. The
HPE enables us freely to vary the initial state pseudoscalar mass
at constant $v.v'$, and we investigate the $1/m_Q$
corrections to the heavy-quark limit.
\end{abstract}

\maketitle

\section{Simulation Details}

We work on a $16^3 \times 48$ lattice at $\beta = 6.0$ in the quenched
approximation. We have analysed a total of 36 configurations, each separated
by 1200 Hybrid Over-Relaxed sweeps (OR and
Cabibbo-Marinari pseudo-heatbath updates in the ratio 5:1)
after allowing 10800 sweeps for thermalisation from a hot start. We use the
$O(a)$-improved Sheikholeslami-Wohlert (SW) fermion action \cite{SWaction} with
the coefficient of the clover term set to its tree-level value (\ie\ $c =
1$). Perturbative $O(a)$-improvement of the operators is achieved by
rotating the quark propagators \cite{UKQCDprocedure}. Light-quark propagators
have been calculated at three
hopping parameter values, \kappalight\ = 0.1432, 0.1440, 0.1445, and the
relevant results from a light hadron analysis are summarised in
table~\ref{tab:lightlight}. The chiral limit corresponds to
\kappacrit = 0.14556(6), and the strange quark to
\kappastrange = 0.1437(5).
These results are consistent with
those recently obtained by the APE collaboration\cite{APE}.
\begin{table}[t]
\caption{The light-light meson spectrum (lattice units).
\label{tab:lightlight} }
\begin{tabular}{cccc} \hline
\kappalight & $m_\pi$ & $m_\rho$ & $f_\pi/Z_A$ \\ \hline 
0.1432 & 0.386\er{4}{4} & 0.51\er{2}{1} & 0.088\er{2}{3} \\ 
0.1440 & 0.311\er{6}{5} & 0.47\er{3}{2} & 0.080\er{2}{4} \\ 
0.1445 & 0.257\er{5}{6} & 0.43\er{6}{3} & 0.075\er{2}{5} \\ \hline 
\kappacrit & --- & 0.38\er{5}{4} & 0.065\er{2}{6}
\\ \hline
\end{tabular}
\end{table}

\section{Pseudoscalar Decay Constants}

\begin{figure}[t]
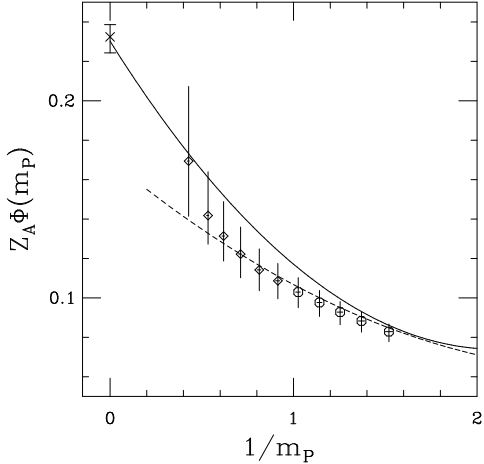

\vspace{-5mm}
\ewxy{fball.ps}{70mm}
\caption{$Z_A \Phi(m_P)$ vs. $1/m_P$ in lattice units.\label{fig:frootm}}
\vspace{-5mm}
\end{figure}

Two-point functions are computed using the Hopping Parameter Expansion (HPE)
\cite{HPE}, evaluated to $200^{\rm th}$ order to ensure that all heavy-light
meson correlators are converged for \kappaheavy\ $<$ 0.133 (significantly
lighter than \kappacharm $\simeq$ 0.125). We use extended operators,
produced by the gauge-covariant Jacobi smearing algorithm,
with an \rms\ smearing radius of four lattice units.
The heavy-light
pseudoscalar decay constants are extracted by analysing the ratio of
the local-smeared axial-pseudoscalar and smeared-smeared
pseudoscalar-pseudoscalar correlators and then extrapolating (interpolating)
the light quark to \kappacrit (\kappastrange) \cite{HeavyDecay}.
Guided by the Heavy Quark Effective Theory (HQET),
we extrapolate (interpolate) the results to $m_B$ ($m_D$) by fitting
\begin{equation}
\Phi(m_P) \equiv Z_A^{-1}f_P\sqrt{m_P}
         \left(\alpha_s(m_P)/\alpha_s(m_B)\right)^{2/\beta_0}
\label{eqn:phi}
\end{equation}
to a quadratic function of $1/m_P$.
This fit is the dotted curve in fig.~\ref{fig:frootm}, where only the
five lightest values of $m_P$ (circles) have been included.

We also compute
two-point functions with a static (\ie\ infinitely massive) heavy quark
using the same smearing radius. A similar analysis to that performed for
propagating heavy quarks yields $Z_L^{\rm stat}$ = 0.211(7),
consistent with APE \cite{APE}.
The corresponding value of $f_B^{\rm stat}$ is 
shown as a cross at $1/m_P$ = 0 in fig.~\ref{fig:frootm}.
We use the perturbative
expressions for $Z_A$,
evaluated with a boosted coupling, yielding $Z_A$ = 0.96 and $Z_A^{\rm
stat}$ = 0.78.

The results are summarised in table~\ref{tab:fP},
where we take the scale from $m_{\rho}$. The first error is
statistical, the second the systematic error from the scale
uncertainty.
Having computed both propagating and static results on the same
configurations, we can perform a fully correlated fit of
eqn.~(\ref{eqn:phi}) to a quadratic function of $1/m_P$,
including the static point.
This fit is the solid curve in fig.~\ref{fig:frootm},
and has a $\chi^2/{\rm dof}$ of 1.5 showing that the 
data are consistent.
Extrapolating to $m_B$ in this fashion raises the value of $f_B$ by 25\%.

\begin{table}[b]
\caption{Pseudoscalar Decay Constants.
\label{tab:fP} }
\begin{tabular}{ll} \hline
$f_D$ & 199\er{14}{15}\er{27}{19}~~MeV \\
$f_B$ & 176\er{25}{24}\er{33}{15}~~MeV \\
$f_{D_s}$ & 225\er{16}{17}\er{31}{21}~~MeV \\
$f_{B_s}$ & 206\er{29}{28}\er{39}{18}~~MeV \\
$f_B^{\rm stat}$ & 286\er{8}{10}\er{67}{42}~~MeV\\ \hline
\end{tabular}
\end{table}

\section{The Isgur-Wise Function}
The matrix element relevant to the semileptonic pseudoscalar to pseudoscalar
decay $P \rightarrow P'$ has the general decomposition
\begin{eqnarray}
\langle P' | V_\mu | P \rangle = \sqrt{m_{P'} m_P} ( & h_+(v.v')(v+v')^\mu + \nonumber\\
 & h_-(v.v') (v-v')^\mu )
\label{eqn:formfactors}
\end{eqnarray}
In the HQET, $h_+$ is the universal Isgur-Wise function, $\xi$, and $h_-$
is zero \cite{iw}. The perturbative corrections to these simple relations,
up to order $\alpha_s^2 z^2$ (where $z = m_{P'}/m_P$),
have been calculated by Neubert \cite{Neubert}. The HQET
symmetry-breaking corrections have not been calculated, but $h_+$ is
protected by Luke's theorem \cite{luke}
so the corrections will be of order $1/m_P^2$ for $h_+$ at zero recoil.

\begin{figure}[t]
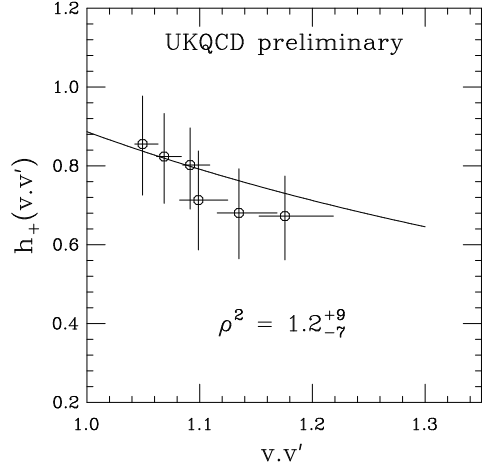

\vspace{-5mm}
\ewxy{hplus_free_rad_10_12.ps}{70mm}
\caption{The Isgur-Wise function.\label{fig:IsgurWise}}
\vspace{-5mm}
\end{figure}

We calculate three-point functions at one value of the
light-quark mass, \kappalight\ = 0.1440, three values
of the the final state
heavy-quark (charm) mass, \kappaheavyprime\ = 0.129, 0.125 and 0.120, and
use the HPE for the
initial state heavy quark (bottom).
The $B$ meson is placed at rest, and the $D$ meson
can have momentum $(8/\pi) |{\bf p}|$ = 0, 1 or $\sqrt{2}$.

To extract $\xi(v.v')$ we set \kappaheavy\ = \kappaheavyprime\ in the HPE and
compute the elastic matrix element $\langle D'| V_4 | D \rangle$ \cite{soni}.
We obtain $h_+$ from eqn.~(\ref{eqn:formfactors}), apply
radiative corrections, then fit to the
BSW ansatz $\xi_\rho(v.v')$ to obtain the slope parameter
$\rho^2=-\xi_\rho'(1)$ \cite{jns} (see fig.~\ref{fig:IsgurWise}).
This yields $\rho^2$=1.2\er{9}{7}, consistent with other determinations.

\begin{figure}[t]
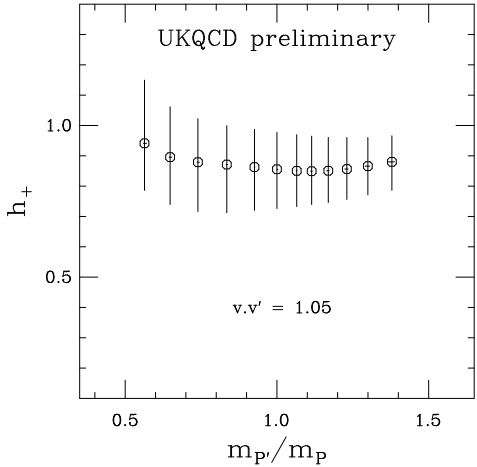

\vspace{-5mm}
\ewxy{f0rad_D_2000.ps}{70mm}
\caption{$h_+$ with radiative corrections.\label{fig:fplus}}
\vspace{-5mm}
\end{figure}

We can exploit the fact that $v.v' = E_{P'}/m_{P'}$ is independent of
\kappaheavy\ to explore the $1/m_P$ corrections to the HQET using the
HPE. For \kappaheavy\ $\neq$ \kappaheavyprime, we simultaneously fit
the matrix elements
of the spatial and temporal components of the vector current to extract both
$h_+$ and $h_-$. In fig.~\ref{fig:fplus} we plot $h_+$
(radiatively corrected)
against $z$ at fixed $v.v'$ and
fixed $m_{P'}$ (\kappaheavyprime\ =
0.120). Since $h_+$ is protected by Luke's theorem, we
would expect the $1/m_P$ corrections to be small close to zero recoil.
This is confirmed by our results, $h_+$ being roughly constant over a
large mass range.

\begin{figure}[t]
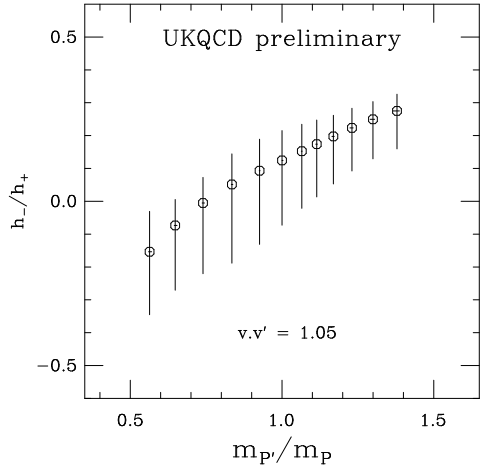

\vspace{-5mm}
\ewxy{f1byf0_D_2000.ps}{70mm}
\caption{$h_-/h_+$ without radiative corrections.\label{fig:hminus}}
\vspace{-5mm}
\end{figure}

In fig.~\ref{fig:hminus} we plot $h_-/h_+$ as a function of $z$.
In contrast to $h_+$, we see that there is a very strong
dependence of $h_-$ on the pseudoscalar mass ratio.
The observed slope is by no means
accounted for purely by the radiative corrections, so we interpret
fig.~\ref{fig:hminus} as evidence for the existence of large $1/m_P$
contributions. This is to be expected as Luke's theorem does
not apply to $h_-$. Due to current
conservation, $h_-$ should vanish for elastic scattering, and
it is reassuring that it is indeed consistent with zero 
when $m_P = m_{P'}$.

\section{Conclusions}
We have shown that the use of rotated operators with the SW action 
gives a sensible normalisation for propagating heavy-quark fields,
yielding values for pseudoscalar decay constants that are consistent
between propagating and static heavy quarks. For semileptonic decays we
have computed the Isgur-Wise function assuming that the HQET is valid,
and find a slope parameter $\rho^2$=1.2\er{9}{7}.
We have shown that the order $1/m_P$ corrections to the HQET predictions
for the form factors are small for $h_+$ but large for $h_-$. These
results are consistent with Luke's theorem.

\section{Acknowledgements}
This work was carried out on the Connection Machine 200 at the University of
Edinburgh; support for the procurement of this system from SERC,
Scottish Enterprise and the Information Systems Committee
of the UFC is acknowledged.
The author acknowledges the support of SERC through grant GR/H01069
and the award of a Postdoctoral Fellowship.

\end{document}